\documentstyle[preprint,aps]{revtex}	
\begin{document}
\tighten
\newcommand{\Jpsi}{J\!/\!\psi}
\newcommand{\mpsi}{m_\psi}
\newcommand{\lms}{\mbox{$\Lambda_{\overline{\mbox{\tiny MS}}}$}}
\def\Lms#1{\Lambda_{\overline{MS}}^{(#1)}}
\def\npb#1#2#3{{\it Nucl. Phys. }{\bf B #1} (#2) #3}
\def\plb#1#2#3{{\it Phys. Lett. }{\bf B #1} (#2) #3}
\def\prd#1#2#3{{\it Phys. Rev. }{\bf D #1} (#2) #3}
\def\prl#1#2#3{{\it Phys. Rev. Lett. }{\bf #1} (#2) #3}
\def\prc#1#2#3{{\it Phys. Reports }{\bf C #1} (#2) #3}
\def\pr#1#2#3{{\it Phys. Reports }{\bf #1} (#2) #3}
\def\zpc#1#2#3{{\it Z. Phys. }{\bf C #1} (#2) #3}
\def\ptp#1#2#3{{\it Prog.~Theor.~Phys.~}{\bf #1} (#2) #3}
\def\nca#1#2#3{{\it Nouvo~Cim.~}{\bf A #1} (#2) #3}
\def\thebibliography#1{{\bf{References}}\list
 {[\arabic{enumi}]}{\settowidth\labelwidth{[#1]}\leftmargin\labelwidth
   \advance\leftmargin\labelsep
   \usecounter{enumi}}
   \def\newblock{\hskip .11em plus .33em minus -.07em}
   \sloppy
   \sfcode`\.=1000\relax}
  \let\endthebibliography=\endlist
\font\fortssbx=cmssbx10 scaled \magstep2
\hbox to \hsize{
\includegraphics{uwlogo.ps}
\hskip.5in \raise.1in\hbox{\fortssbx University of Wisconsin - Madison}
\hfill$\vtop{\hbox{\bf MADPH--94--860}
                \hbox{September 1994}}$ }
\vspace{1cm}
\begin{center}{\bf
$\Jpsi$ Decay Lepton Distribution
in Hadronic Collisions }\end{center}

\begin{center}{E. Mirkes}\end{center}
\begin{center}{
\it Physics Department, University of Wisconsin, Madison WI 53706, USA}
\end{center}
\begin{center}{C.S. Kim }\end{center}
\begin{center}{
\it  Department  of Physics, Yonsei University, Seoul 120--749, KOREA}
\end{center}
\begin{abstract}
We propose the measurement of the
decay angular distribution
of leptons from $\Jpsi$'s produced at high transverse momentum
balanced by a photon [or gluon]
in hadronic collisions.
The polar and azimuthal angular distribution
are calculated in the color singlet model (CSM).
It is shown that the general structure of the decay lepton distribution
is controlled by four invariant structure functions, which are functions
of the transverse momentum and the rapidity of the $\Jpsi$.
We found that two of these structure functions
[the longitudinal and transverse interference structure functions]
are identical in the CSM.
We present analytical and numerical results in the  Collins-Soper
and in the Gottfried-Jackson frame.
\end{abstract}

\newpage
The measurement of the angular distribution of lepton's
from $\Jpsi$'s provides a detailed test of the production
and decay mechansim of the $c\bar{c}$ bound state.
So far there has been intensive experimental studies of the $\Jpsi$
production rate and transverse momentum distributions
in hadronic collisions both
at UA1 \cite{ua1} and CDF \cite{cdf}.
However, the observed $\Jpsi$ rate was found to be markedly higher
than the predicted one's in \cite{theo}, where
in addition to the direct charmonium production also contributions from
the $\chi$ and the production resulting from $B$ decays were
taken into account.
It has recently been pointed out in \cite{eric} that at large
transverse momentum of the $\Jpsi$  an additional production mechanism
comes from fragmentation contributions of a gluon or a charm quark
into charmonium states.

In this letter
we restrict ourselves  to the study of $\Jpsi$ produced
in association with a photon \cite{kim}. This has several advantages.
First of all the experimental signature of a $\Jpsi$ [decaying into
an $e^-e^+$ or $\mu^-\mu^+$ pair] and a $\gamma$ with balancing
transverse momentum is a very clean final state.
Second, the  fragmentation  contributions \cite{roy}
from radiative $\chi$ decays\footnote{
The radiative $\chi_J$ decays can produce
$J/\psi$ at both low and high $p_{_T}$, but the photon produced will be
soft [$E \sim {\cal O}(400\;{\rm MeV})$].}
to $\Jpsi + \gamma$ production are small and the dominant subprocess
contributing to a  $\Jpsi+\gamma$ final state is
the gluon gluon fusion process.

For $\Jpsi$'s produced with
transverse momentum $p_{_T}$ [balanced by the additional photon] one
can define an event plane spanned by the beam and the
$\Jpsi$ momentum direction which provides
a reference plane for a detailed study of angular correlations.
The decay lepton distribution in $\Jpsi\rightarrow l^- l^+$
in the $\Jpsi$ rest frame is determined by the polarization of the $\Jpsi$.
Therefore, the study of the angular distribution can be used as an analyzer
of the $\Jpsi$-polarization.
It is thus possible to test the underlying $\Jpsi$-production dynamics
[in our case the color singlet model (CSM) \cite{csm}]
in much more detail than is possible by rate measurements alone
\cite{volker}.

For definiteness
we consider the angular distribution of the leptons coming from the
leptonic decay of $\Jpsi$'s produced with non-zero transverse
momentum in association with a photon
in high energy proton-antiproton collisions:
\begin{equation}
p(P_{1}) + \bar{p}(P_{2}) \rightarrow \Jpsi(P) +\gamma(k) +X
\rightarrow l^-(l)+l^+(l') + \gamma(k)+ X \>,
\end{equation}
where the quantities in the parentheses denote the four-momenta.
In  leading order of perturbative QCD [$O(\alpha_s^2)$],
 $\Jpsi+\gamma$  can only be produced in $gg$
fusion:
  \begin{equation}
g(p_{_1})+g(p_{_2})\rightarrow \Jpsi(P)+ \gamma(k) +X\>.
\label{partonic}
  \end{equation}
Denoting hadron level and parton
level quantities by upper and lower case characters, respectively,
the hadron and parton level Mandelstam variables are  defined by
\begin{equation}
S = (P_1 + P_2)^2 \>, \qquad T = (P_1 - P)^2 \>, \qquad U = (P_2 - P)^2 \>,
\end{equation}
and
  \begin{equation}
  \begin{array}{lclcl}
\label{kleinmandeldef}
s&=&(p_{_1}+p_{_2})^{2}&=&x_{1}x_{2}S \>,\\[2mm]
t&=&(p_{_1}-P)^{2}    &=&x_{1}(T-P^2)+P^2\>,\\[2mm]
u&=&(p_{_2}-P)^{2}    &=&x_{2}(U-P^2)+P^2\>,
  \end{array}
\end{equation}
The angular distribution of the leptons
from the $\Jpsi$ has the general form
\begin{eqnarray}
\frac {d\sigma}{d p_{_T}^{2}\,dy\, d\cos\theta \,d\phi}
&=&  \frac{3}{16\pi}\,
\frac{d\sigma^{U+L}}{ d p_{_T}^{2}\,dy}\,\,
           \,\left[  (1+\cos^{2}\theta) \nonumber
               +\,\, \frac{1}{2}A_{0} \,\,\, (1-3\cos^{2}\theta)
 \right.\\[2mm]
&&   \left.
\hspace{1.5cm} +  \,\,   A_{1}  \,\,\,\sin 2\theta \cos\phi \,\,
\,\, + \,\,   \frac{1}{2}A_{2}
\,\,\,\sin^{2}\theta\cos 2\phi    \right]\>.
  \label{ang}
\end{eqnarray}
The angles $\theta$ and
$\phi$  in Eq. (\ref{ang})
are the polar and azimuthal
decay angles of the leptons  in the $\Jpsi$ rest frame with respect to
a coordinate system  described below and $y$ ($p_T$) denotes the rapidity
(transverse momentum) of the $\Jpsi$ in the laboratory frame.
Note that the unpolarized differential production cross section  denoted
by ${\sigma}^{U+L}$ is factored out from the
r.h.s of Eq.~(\ref{ang}).
The angular coefficients $A_i$
characterize the polarization of the $\Jpsi$ (see below).
They are dependent on the choice of the $z$ axis in the $\Jpsi$ rest frame
and are defined by the following ratios of helicity cross sections
$\left(d\sigma^{U+L,L,T,I}
\equiv\frac{d{\sigma}^{U+L,L,T,I}}{dp_{_T}^2\,dy}\right)$
\begin{equation}
A_{0}=\frac{2\,\, d\sigma^{L}}{d\sigma^{U+L}}\>,\hspace{1cm}
A_{1}=\frac{2\sqrt{2}\,\, d\sigma^{I}}{d\sigma^{U+L}}\>,\hspace{1cm}
A_{2}=\frac{4 \,\,d\sigma^{T}}{d\sigma^{U+L}}\>,\hspace{1cm}
\label{aintr}
\end{equation}
The hadronic  helicity cross sections
$\frac{d{\sigma}^{U+L,L,T,I}}{dp_{_T}^2\,dy}$
are obtained by convoluting the partonic helicity  cross sections
$\frac{{s}\,d{\hat{\sigma}^{U+L,L,T,I}}}{ d{t}\,d{u}}$
with the parton densities. One has
\begin{equation}
\frac{d\sigma^{U+L,L,T,I}}{ dp_{_T}^{2}\,dy } =
\int\,\,dx_{1}dx_{2}
g^{h_{1}}(x_{1},\mu_F^{2})
g^{h_{2}}(x_{2},\mu_F^{2})
\,
\frac{{s}\,d{\hat{\sigma}^{U+L,L,T,I}}}{ d{t}\,d{u}}
\label{wqalphahad}\>.
\end{equation}
Each of the partonic  helicity cross sections is calculated
in the CSM.
The unpolarized differential production cross section is denoted
by $\hat{\sigma}^{U+L}$ whereas $\hat{\sigma}^{L,T,I}$ characterize the
polarization of the $\Jpsi$, i.e.
the  cross section for the longitudinal polarized $\Jpsi$'s is denoted
by $\hat{\sigma}^L$, the transverse-longitudinal
interference by $\hat{\sigma}^I$,
and the transverse interference
by $\hat{\sigma}^T$
(all with respect to the $z$-axis of the chosen $\Jpsi$ rest frame).
The results for the helicity cross sections $\hat{\sigma}^{U+L,L,T,I}$
are dependent of the choice of the
$z$ axis in the rest frame of the $\Jpsi$. We present
explicit results for the
Collins-Soper (CS) and the Gottfried-Jackson (GJ) frame:

In the CS frame \cite{cs} the $z$-axis
bisects the angle between ${\vec{P}_{1}}$ and ${-\vec{P}_{2}}$
\begin{equation}
    \begin{array}{ll}
CS: \,\,\,& {\vec{P}_{1}} =
\,\,\, E_{1}\,\, (\sin\gamma_{CS},0,\cos\gamma_{CS})\>,   \\[1mm]
          & {\vec{P}_{2}} =
\,\,\,     E_{2}\,\,(\sin\gamma_{CS},0,-\cos\gamma_{CS})\>,
          \end{array}
\label{csdef}
\end{equation}
with
%
$
\cos\gamma_{CS}=
\sqrt{\frac{\mpsi^2 S}{(T-\mpsi^2)(U-\mpsi^2)}}
=\sqrt{\frac{\mpsi^2}{\mpsi^2+p_{_T}^2}}\>,
%
%
\hspace{3mm}
\sin\gamma_{CS}=-\sqrt{1-\cos^2\gamma_{CS}}\>.
$
%

In the GJ frame
(also known as $t$-channel helicity frame)
the $z$-axis is chosen parallel to the beam axis
\begin{equation}
          \begin{array}{ll}
GJ: \,\,\,& {\vec{P}_{1}} = E_{1} \,\,\,
                            (0,0,1)\>,\\[1mm]
          & {\vec{P}_{2}} = E_{2}\,\,\,
                            (\sin\gamma_{GJ},0,\cos\gamma_{GJ})\>,
          \end{array}
\label{gjdef}
\end{equation}
with
%

$
\cos\gamma_{GJ}=
 {1-\frac{2\mpsi^2 S}{(T-\mpsi^2)(U-\mpsi^2)}}
=\frac{p_{_T}^2-\mpsi^2}{p_{_T}^2+\mpsi^2}\>,
%
%
\hspace{3mm}
\sin\gamma_{GJ}=-\sqrt{1-\cos^2\gamma_{GJ}}\>.
$
%
The beam and target energies in the rest frame are
$E_{1}=(\mpsi^2-T)/(2\mpsi), \,\,\,\, E_{2}=(\mpsi^2-U)/(2\mpsi)~.$

The partonic helicity cross sections in Eq. (\ref{wqalphahad})
are calculated applying the technique described in \cite{npb}
to the CSM [details of the calculation will be presented elsewhere].
They are given by
[$R(0)$ denotes the radial wave function of the bound state]:
\begin{equation}
\frac{s\hat{\sigma}^{\alpha}_{CS, GJ}}{dt du}
=\frac{16\pi \alpha\alpha_s^2 \mpsi}{27 s }
|R(0)|^2\,\,H^{\alpha}_{CS, GJ}\,
\delta(s+t+u-\mpsi^2)\>,
\label{partondef}
\end{equation}
with
\begin{eqnarray}
H_{CS}^{U+L}&=&
       \frac{s^2}{(t-\mpsi^2)^2(u-\mpsi^2)^2}
     + \frac{t^2}{(u-\mpsi^2)^2(s-\mpsi^2)^2}
     + \frac{u^2}{(s-\mpsi^2)^2(t-\mpsi^2)^2}\>,\hspace{1cm}\label{firstcs}
\\[2mm]
H_{CS}^{L}&=&
             \frac{ ut}{2(t-\mpsi^2)(u-\mpsi^2)}\left(
         \frac{4\mpsi^2s^3}{(s-\mpsi^2)^2(t-\mpsi^2)^2(u-\mpsi^2)^2}
          + H^{U+L}_{CS}\right)\>,\\[2mm]
H_{CS}^{T}&=&\frac{1}{2} H_{cs}^{L}\>,\\[2mm]
H_{CS}^{I}&=&
\frac{ \mpsi\sqrt{sut}\,\,\,s(s^2-ut)(t-u)}{
        \sqrt{2}(s-\mpsi^2)^2(t-\mpsi^2)^3(u-\mpsi^2)^3}\>,
\label{lastcs}
\end{eqnarray}
and for the GJ frame
\begin{eqnarray}
H_{GJ}^{U+L}&=& H_{CS}^{U+L}\>,\hspace{12cm}\label{firstgj}\\[2mm]
H_{GJ}^{L}&=&
 \frac{2\mpsi^2 s t u
(s^2+u^2)}{(s-\mpsi^2)^2(t-\mpsi^2)^4(u-\mpsi^2)^2}\>,\\[2mm]
H_{GJ}^{T}&=&\frac{1}{2} H_{GJ}^{L}\>,\\[2mm]
H_{GJ}^{I}&=&
-\frac{\mpsi\sqrt{ s t u }\,(s-u)[s^2(u-t)+u^2(s-t)]}{
        \sqrt{2}(s-\mpsi^2)^2(t-\mpsi^2)^4(u-\mpsi^2)^2}\>.
 \label{lastgj}
\end{eqnarray}
As mentioned before, $H^{U+L}$ denotes the matrix element contribution
for the production rate and was first calculated in
the first paper of \cite{csm}.
All other matrix elements correspond to the production of
{\it polarized} $\Jpsi$'s and are given here for the first time.
Replacing  $16\alpha$  by $15\alpha_s$ in Eq.~(\ref{partondef}),
the results in
Eqs.~(\ref{partondef}-\ref{lastgj}) are also valid for $\Jpsi+g$
production within the CSM.

We will now present numerical results for $\Jpsi+\gamma$ production  at
the Tevatron collider center of mass energy [$\sqrt{S}=1.8$ TeV] including
the decay  $\Jpsi\rightarrow 
\mu^-\mu^+$.
All results are obtained using the  gluon density
parametrization from  GRV \cite{grv}
with $\Lms{4}=200$ MeV
and the one-loop formula for $\alpha_s$ with 4 active flavours.
If not stated otherwise,
we identify the renormalization scale $\mu_R^2$
and the factorization scale $\mu_F^2$ and set
them equal to
$\mu^2=\mu_F^2=\mu_R^2=(\mpsi^2+p_{_T}^2(\Jpsi))$.
The value for the bound state wave function at the origin $|R(0)|^2$
is determined from the leptonic decay width of $\Jpsi$:
$\Gamma (\Jpsi \rightarrow e^- e^+) = 4.72 $ keV, therefore
$|R(0)|^2 = 0.48$  GeV$^3$.

In Figs.~1 and 2 we show
numerical results for the coefficients $A_i$ in Eq.~(\ref{ang})
as a function of $p_T(\Jpsi)$ and $y(\Jpsi)$
 in the CS [Figs.~1,2(a)] and GJ [Figs.~1,2(b)] frame.
One observes that the coefficients are  strongly dependent
on $p_T(\Jpsi)$ both in the CS and GJ frame.
As  mentioned before, the coefficients $A_0$ and $A_2$ are exactly equal
in lowest order in both lepton pair rest frames.
The angular coefficient $A_1$ is zero in the CS frame
for all values of $p_{_T}(\Jpsi)$.
The reason is that the matrix element
for $A_1$ is antisymmetric in $u$ and $t$ [see Eqs. (\ref{lastcs},\ref{aintr})]
and therefore in $x_1$ and $x_2$, whereas
the product of the gluon distributions is symmetric under the interchange
of $x_1$ and $x_2$.
As a consequence, the rapidity distribution for $A_1$ in the
CS frame [Fig.~2(a)] is also antisymmetric around $y=0$.
However, this is  different   for the GJ frame [see Eq. (\ref{lastgj})].
All  coefficients $A_i$ vanish in the limit
$p_{_T}(\Jpsi)\rightarrow 0$, which can be
directly seen from our analytical expressions in Eqs.
(\ref{firstcs}-\ref{lastgj},\ref{aintr}).

A similar relation $A_0^{DY}=A_2^{DY}$
was found in LO [$O(\alpha_s)$] in the Drell-Yan process \cite{tung}
$p+\bar{p}\rightarrow V + X \rightarrow l^+l'+X$, where $V$ denotes a
gauge boson produced at high $p_T$.
In \cite{npb}, the complete NLO corrections to the coefficients
$A^{DY}_i$ are calculated and  the corrections
are found to be fairly small for the ratio's $A_i$ of the corresponding
helicity cross section.
We expect also here, that the  LO results
for the ratio's $A_i$ in $\Jpsi$ production
are almost not affected by higher order
QCD corrections. Note also, that the coefficients $A_i$
are not dependent on the bound state wave function.
The measurement of these coefficients
would be a sensitive  test of the production mechanism of $\Jpsi$'s.

To summarize:
hadronic $\Jpsi+\gamma$ production has been evaluated in the
nonrelativistic bound state model.
In leading order this final state can only be produced by gluon fusion.
Analytical formula for
the decay lepton distributions
in terms of four  structure functions are presented.
The angular distribution
in the $\Jpsi$ rest frame is  determined by the polarization of the $\Jpsi$.
\\[1cm]
\bigskip
\noindent{\bf Acknowledgements}
\medskip

The work of EM is supported in part by the U.S. Department of Energy under
contract Nos. DE-AC02-76ER00881 and DE-FG03-91ER40674,  by Texas
National Research Laboratory Grant No.~RGFY93-330, and by the
University of Wisconsin Research Committee with funds granted by the
Wisconsin Alumni Research Foundation.
The work of CSK is supported in part by the Korean Science and Engineering
Foundation, in part by Non-Direct-Research-Fund, Korea Research Foundation
1993, in part by the Center for Theoretical Physics, Seoul National University,
and in
part by the Basic Science Research Institute Program, Ministry of Education,
1994, Project No. BSRI-94-2425.

\noindent
{\bf Figure captions}
\begin{itemize}
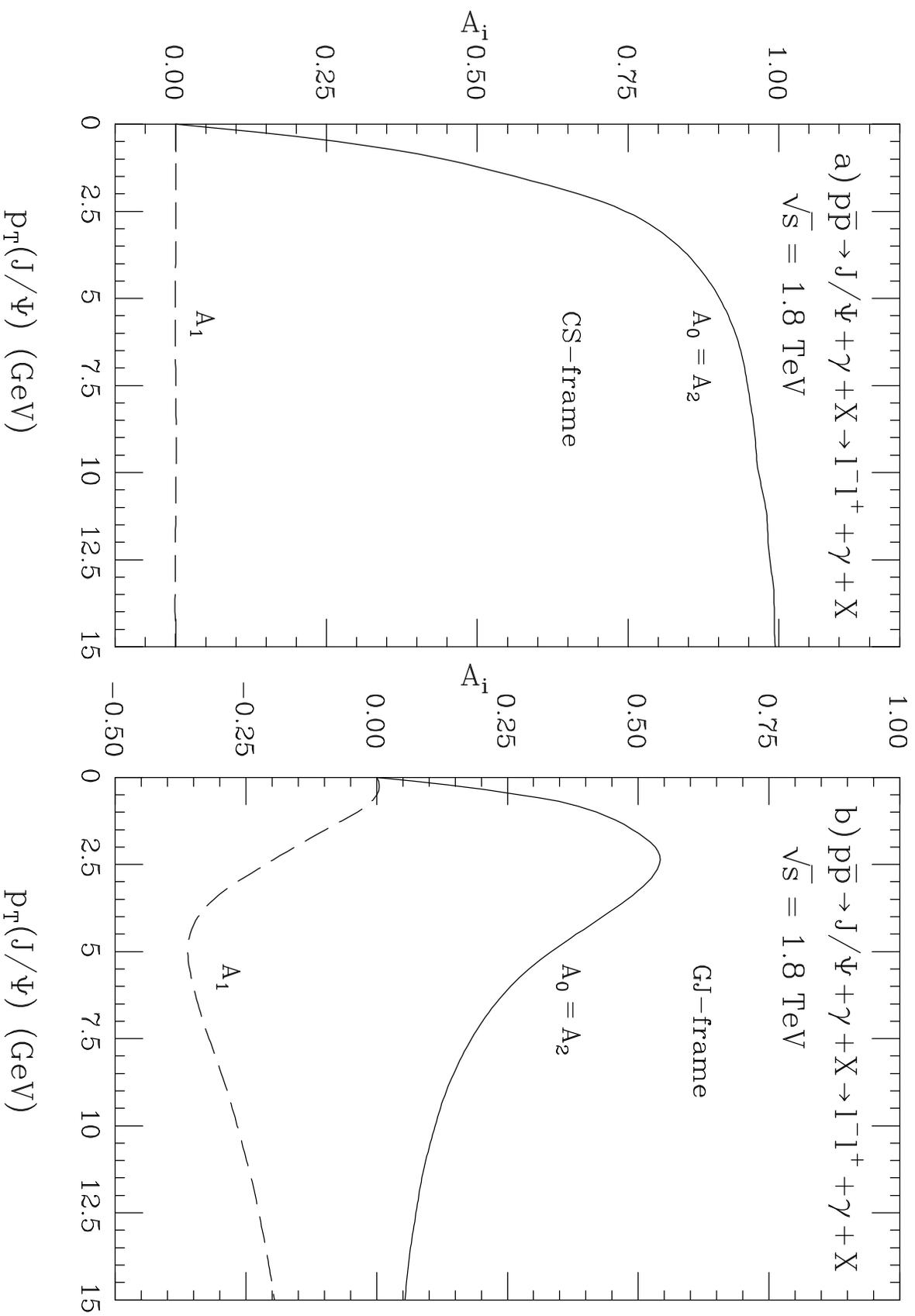
\item[{\bf Fig. 1}]
Angular coefficients $A_0, A_1$ and $A_2$ for $\Jpsi+\gamma$
production and decay in the
CS frame (a) and GJ frame (b)
as a function of the $\Jpsi$ transverse momentum
at $\sqrt{S}=1.8$ TeV.
No cuts have been applied.
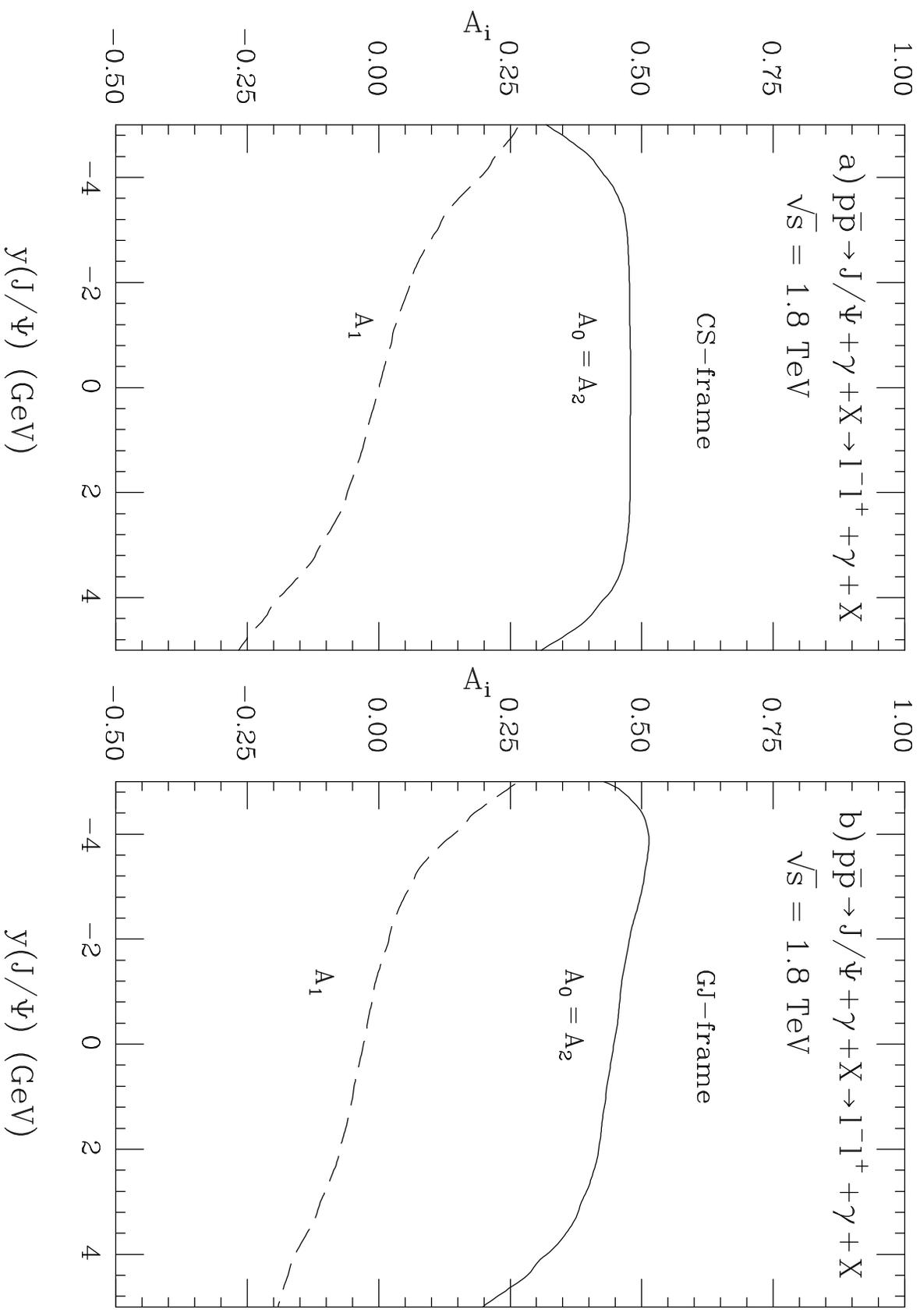
\item[{\bf Fig. 2}]
Same as Fig.~1 for the $y(\Jpsi)$ distribution.
\end{itemize}
%

\raggedright

\end{document}